\documentclass[prd,eqsecnum,notitlepage,%linenumbers,%showpacs,%preprint,
nofootinbib,superscriptaddress]
{revtex4-1}
\global\arraycolsep=2pt
\usepackage{stmaryrd}
\usepackage{amsmath}
\usepackage{amssymb}
\usepackage{graphicx}
\usepackage{textcomp}
\usepackage{calrsfs}
\usepackage{yfonts}
\usepackage{bm}
\usepackage{color}

\usepackage{caption}
\usepackage{subcaption}

\newcommand{\ben}{\begin{equation*}} 
\newcommand{\een}{\end{equation*}}
\newcommand{\bean}{\begin{eqnarray*}}
\newcommand{\eean}{\end{eqnarray*}}

\newcommand{\nn}{\nonumber}
\newcommand{\be}{\begin{equation}} 
\newcommand{\ee}{\end{equation}}
\newcommand{\bea}{\begin{eqnarray}}
\newcommand{\eea}{\end{eqnarray}}

\DeclareMathOperator{\Tr}{Tr}

\begin{document}

\title{Negativity of the Casimir self-entropy in spherical geometries}

% Author Orchid ID: enter ID or remove command
%\newcommand{\orcidauthorA}{0000-0001-5973-8980} % Add \orcidA{} behind the author's name
%\newcommand{\orcidauthorB}{0000-0002-7148-0609} % Add \orcidB{} behind the author's name
%\newcommand{\orcidauthorC}{0000-0003-0812-0421} % Add \orcidC{} behind the author's name

\author{Yang Li}
\email{leon@ncu.edu.cn}
\affiliation{Department of Physics,
Nanchang University, Nanchang 330031, China}

\author{Kimball A. Milton}
  \email{kmilton@ou.edu}
  \affiliation{H. L. Dodge Department of Physics and Astronomy,
University of Oklahoma, Norman, OK 73019, USA}

\author{Prachi Parashar}
\email{Prachi.Parashar@jalc.edu}
\affiliation{John A. Logan College, Carterville, IL 62918 USA}

\author{Lujun Hong}
\affiliation{Institute of Space Science and Technology, Nanchang University, 
Nanchang, 330031, China}

% Authors, for the paper (add full first names)
%\Author{Yang Li $^{1,\dagger}$\orcidA{}, 
%Kimball A. Milton $^{2,\dagger,*}$\orcidB{},  
%Prachi Parashar $^{3,\dagger}$\orcidC{},
%and Lujun Hong $^{4}$}

% Authors, for metadata in PDF
%\AuthorNames{Yang Li, Kimball A. Milton,  and Prachi Parashar}

% Affiliations / Addresses (Add [1] after \address if there is only one affiliation.)
%\address{%
%$^{1}$ \quad 
%$^{2}$ \quad Homer L. Dodge Department of Physics and Astronomy,
%University of Oklahoma, Norman, OK 73019 USA; kmilton@ou.edu\\
%$^{3}$\quad 

%$^4$\quad
%}

% Contact information of the corresponding author
%\corres{Correspondence: kmilton@ou.edu%; Tel.: (optional; include country code; if there are multiple corresponding authors, add author initials) +xx-xxxx-xxx-xxxx (F.L.)

% Current address and/or shared authorship
%\firstnote{Current address: Affiliation 3} 
%\firstnote{These authors contributed equally to this work.}
% The commands \thirdnote{} till \eighthnote{} are available for further notes

%\simplesumm{} % Simple summary

%\conference{} % An extended version of a conference paper

% Abstract (Do not insert blank lines, i.e. \\) 
\begin{abstract}
It has been recognized for some time that even for perfect conductors, 
the interaction Casimir entropy, due to quantum/thermal fluctuations, can be 
negative.  This result was not considered problematic because it was thought 
that the self-entropies of the bodies would cancel this negative interaction
entropy, yielding a 
total entropy that was positive.  In fact, this cancellation seems not to occur.
The positive self-entropy of a perfectly conducting sphere does indeed just 
cancel the negative interaction entropy of a system consisting of a perfectly 
conducting sphere and plate, but a model with weaker coupling in general 
possesses a regime where negative self-entropy appears.  The physical meaning 
of this surprising result remains obscure. In this paper we re-examine these 
issues, using improved physical and mathematical techniques, partly based on
the Abel-Plana formula, and present numerical
results for arbitrary temperatures and couplings, which exhibit the same
remarkable features.
\end{abstract}

\date\today
\maketitle
%%%%%%%%%%%%%%%%%%%%%%%%%%%%%%%%%%%%%%%%%%
%\setcounter{section}{-1} %% Remove this when starting to work on the template.
\section{Introduction}
It is ordinarily expected that entropies of closed systems should be positive.
This follows from the Boltzmann definition in terms of the number of 
microstates $\Omega$, so the entropy is given as $S=k_B\ln\Omega$ ($k_B$ is the
Boltzmann constant).  Quantum-mechanically, in terms of the density operator $\rho$, the
entropy is $S=-k_B\Tr \rho\ln\rho$.  But there are intriguing possibilities of negative
entropy \cite{schrodinger,Cvetic:2001bk,Nojiri:2004pf}.

Here we are considering quantum-fluctuational or Casimir free energies and 
entropies.  For two parallel conducting plates possessing nonzero resistivity, 
the entropy corresponding to the interaction free energy vanishes at zero 
temperature, as required by the Nernst heat theorem (third law of 
thermodynamics). However, for sufficiently low temperatures, compared to the 
inverse of the plate separation, a region of negative interaction entropy 
emerges \cite{njp}.  But the expectation at that time was that the total entropy
must be positive.  Negative Casimir interaction entropies also occurred without
dissipation  between a sphere and a plane 
\cite{bezerra,durand1,durand2,bp}, both perfectly conducting, or between   
two perfectly conducting spheres \cite{lopez1,lopez2}. 
This was systematically explored in the dipole regime \cite{milton,ingold}.

But, indeed, it turned out that the sphere-plane problem was resolved by considering the
self-entropy of the plate and the sphere separately.  The former vanishes in the
perfectly conducting limit, but the latter is just such as to cancel the most 
negative contribution of the interaction entropy \cite{li,milton2}.
The sphere-sphere entropy is then seen to be clearly positive as well.

However,
 going beyond the case of a perfectly conducting sphere has proved to be more subtle.
We carried out a systematic treatment for an imperfectly conducting sphere, 
modeled by a $\delta$-function sphere, or a ``plasma-sphere,'' described by the
 potential $\mathbf{V}=\bm{\varepsilon-1}=\lambda(\bm{1-\hat r\hat r})$, 
(in terms of polar coordinates based on the center of the sphere), where the 
transversality condition is required by Maxwell's equations.  We take the coupling
$\lambda$ to be frequency dependent, according to the plasma model,
 $\lambda=\lambda_0/(\zeta^2a)$, where $\zeta=
-i\omega$ is the Euclidean frequency, and $a$ is the radius of the sphere. The
dimensionless coupling constant $\lambda_0$ is necessarily positive.
In the limit of $\lambda\to\infty$ we recover the entropy for a perfectly conducting 
sphere first obtained by Balian and Duplantier \cite{bd}.  But for sufficiently
weak coupling, even at high temperatures, we found that the entropy could turn
negative \cite{milton3,milton4}.  (The results found there largely agreed with those
found subsequently by Bordag and Kirsten \cite{bordagandkirsten,bordagfs}.)

Since the transverse electric contribution to the entropy is always negative,
and presents no difficulties in its evaluation, 
in this paper we concentrate on the transverse
magnetic free energy, $F_H$. One feature of the analysis here is that
we always subtract an infrared sensitive, but unphysical term, which we only subtracted
in a {\it ad hoc\/} manner in Ref.~\cite{milton3}. The most salient element of
our new treatment, however, is the emphasis on the Abel-Plana formula, and the numerical
computations based upon that formulation.
  In the next section we give the general formulas for this
model, and recast the result in Abel-Plana form, which expresses the 
finite temperature-dependent part of free energy in
terms of a mode sum over the phase of a quantity involving spherical Bessel functions.
Then in Sec.~\ref{weaksec} we specialize to weak coupling, where the mode sum can be
carried out explicitly for the lowest-order term. The result agrees with that found in
Ref.~\cite{milton3}.  The low-temperature limit is considered in Sec.~\ref{lowtsec};
we extract coincident free energies
using both the Euclidean and the (real-frequency) Abel-Plana
formulations.  We briefly review the previous result for high temperatures in 
Sec.~\ref{htempsec}.  Finally, we present general  numerical results in Sec.~\ref{numsec},
which, for coupling and temperature of order unity (in units of $1/a$) 
turn out to be remarkably similar to those found for low temperature. Further numerical
explorations have shown how the analytic asymptotic behaviors are realized.
 Concluding remarks round out the paper.

In this paper with adopt natural units, with $\hbar=c=k_B=1$.

\section{Transverse magnetic free energy of plasma-shell sphere}
We concentrate on the transverse-magnetic (TM) contribution to the free energy
of a $\delta$-sphere, since the transverse electric (TE) part seems unambiguous,
and always yields a negative contribution to the entropy.
As derived in Ref.~\cite{milton3} the TM free energy is given by
\be
F_H=\frac{T}2\sum_{n=-\infty}^\infty e^{in\alpha\tilde\tau}\sum_{l=1}^\infty 
(2l+1) P_l(\cos\delta)\ln\left[1-\lambda_0\frac{\alpha|n|e_l'(\alpha|n|)
s_l'(\alpha|n|)}{\alpha^2n^2+\tilde\mu^2}\right],\label{ftme}
\ee
where $\tilde\tau=\tau/a\to0$ is the dimensionless time-splitting regulator, 
$\delta\to0$ is the angular point-splitting regulator, and $\alpha=2\pi a T$, so
that $n\alpha=a\zeta_n$, where $\zeta_n$ is the Matsubara frequency.  Further we
have inserted an infrared regulator $\tilde\mu=\mu a$, modeled as a photon mass.
Here the modified Ricatti-Bessel functions are
\be
s_l(x)=\sqrt{\frac{\pi x}2}I_{l+1/2}(x),\quad e_l(x)=\sqrt{\frac{2x}\pi}K_{l+1/2}(x).
\ee
We might hope to  eliminate the $\tilde\mu$ regulator dependence, formally, by subtraction
of an unphysical coupling-independent term:
\be
F_H=\frac\alpha{2\pi a}\sum_{n=0}^{\infty}{}'\cos(n\alpha\tilde\tau)
\sum_{l=1}^\infty (2l+1)P_l(\cos\delta)
\left(\ln\left[(n\alpha)^2+\tilde\mu^2
-\lambda_0f_H(l,n\alpha)\right]-\ln \left[(n\alpha)^2+\tilde\mu^2\right]\right),
\label{fh}
\ee
where the prime on the summation sign means that the $n=0$ term is to be counted 
with half weight, and
 we have abbreviated $f_H(l,x)=x e_l'(x)s_l'(x)$.  The subtracted term was
evaluated in Ref.~\cite{milton3}, because $\sum_{l=1}^\infty (2l+1)P_l(\cos\delta)
=-1$:
\be
F_H^{\rm sub}=\frac{T}2\sum_{n=-\infty}^\infty e^{in\alpha\tilde\tau}\ln
\left[n^2\alpha^2
+\tilde\mu^2\right]=-\frac1{2\tau}+T\ln\frac\mu{T}.\label{fhsub}
\ee
We discarded this term as unphysical (it makes no reference to the properties of
the sphere) frequently throughout Ref.~\cite{milton3}, although it was not
done systematically.  Now we propose doing so.  Then we can recast the remainder
of $F_H$ using the Abel-Plana formula, which reads
\be
\sum_{n=0}^{\infty}{}' g(n)=\int_0^\infty dt\, g(t)
+i\int_0^\infty dt\frac{g(it)-g(-it)}{e^{2\pi t}-1}.
\ee
Applied to Eq.~(\ref{fh}) after the omission of the subtracted term (\ref{fhsub}), 
we see that the first integral gives a contribution independent of $T$, which is
the (divergent) zero-temperature TM energy of the sphere \cite{deltasph}.
  We are here only concerned with the temperature-dependent part, which we can rewrite as
\be
\Delta F_H=-\frac1{\pi a}\int_0^\infty\frac{dx}{e^{2\pi x/\alpha}-1}
\sum_{l=1}^\infty (2l+1)\arg[-x^2-\lambda_0f_H(l,ix)].\label{fhap}
\ee
Here, we have dropped the regulators because this expression is finite.

The definition of the argument function is somewhat subtle.  We choose it to 
be defined by the usual arctangent,
\be
\arg(z)=\arctan\left(\frac{\Im z}{\Re z}\right),\quad 
-\frac\pi2<\arctan y\le \frac\pi2,
\ee
which is discontinuous when $\Re z$ passes through zero. 
This choice is necessary in order to have a well-defined limit at zero 
temperature. (See Section~\ref{sec:ap}.)  
It also guarantees that the free energy
vanishes for zero coupling, which would seem an obvious physical requirement.
 Therefore, the argument appearing in Eq.~(\ref{fhap}) is 
\be
\arg[-x^2-\lambda_0f_H(l,ix)]=\arctan\left(\frac{\lambda_0\frac\pi2\mathcal{J}_\nu^2(x)}
{-x^2+\lambda_0\frac\pi2\mathcal{J}_\nu(x)\mathcal{Y}_\nu(x)}\right),\quad \nu=l+\frac12.
\label{argat}
\ee
%with $\mathcal{Y}_\nu$ be defined as in Eq.~(\ref{JH}) in terms of Bessel functions of
%the second kind.  
The functions appearing here are, in terms of ordinary Bessel functions $J_\nu$ and
$Y_\nu$, 
%\be
%f_H(l,ix)=-i\frac\pi2\mathcal{J}_\nu(x)\mathcal{H}_\nu(x),\quad \nu=l+\frac12,
%\ee
%where
\begin{subequations}
\label{JY}
\bea
\mathcal{J}_\nu(x)&=&-\sqrt{\frac{2x}\pi}[xj_l(x)]'=(\nu-1/2)J_\nu(x)-xJ_{\nu-1}(x),\\
\mathcal{Y}_\nu(x)&=&-\sqrt{\frac{2x}\pi}[x y_l(x)]'=
(\nu-1/2)Y_\nu(x)-xY_{\nu-1}(x),
\eea
\end{subequations}
%$H^{(2)}_\nu(x)=J_\nu(x)-iY_\nu(x)$ being the Hankel function of the second kind, and
$j_l$, $y_l$ being the corresponding spherical Bessel functions.

The ultraviolet  convergence of $\Delta F_H$ in Eq.~(\ref{fhap}) in $x$ is assured 
by the exponential
factor, but the convergence in $l$ requires further investigation.  It is easily
checked that 
\be
f_H(l,ix)\sim - \frac\nu2 \quad\mbox{as}\quad l\to\infty,
\ee
so
\be
\arg[-x^2-\lambda_0 f_H(l,ix)]\to 0, \quad \mbox{as}\quad l\to \infty.
\label{arg}
\ee

\section{Weak coupling}
\label{weaksec}
With the above definition of the argument function, we can readily work out the weak coupling
expansion of the free energy.  The leading term in $\lambda_0$ is obtained from the 
first term in the expansion of the arctangent, so
\be
\Delta F_H^{(1)}=\frac{\lambda_0}{\pi a}\int_0^\infty \frac{dx}{x}\frac1{e^{2\pi x/\alpha}-1}
\sum_{l=1}^\infty (2l+1)\left([x j_l(x)]'\right)^2.\label{wc}
\ee
The sum on $l$ can be carried out using the addition theorem for spherical Bessel functions 
\be
\sum_{l=0}^\infty (2l+1) j_l(x)j_l(y)=\frac{\sin(x-y)}{x-y}.
\ee
Then, the $l$ sum in Eq.~(\ref{wc}) is
\be
\lim_{y\to x}\frac{\partial}{\partial x}\frac\partial{\partial y}xy\left[\frac{\sin (x-y)}{x-y}
-j_0(x)j_0(y)\right]=1+\frac{x^2}3-\cos^2x,
\ee
since $j_0(x)=\frac{\sin x}x$.
This yields the same result found in Ref.~\cite{milton3}, Eq.~(5.34),
\be
\Delta F_H^{(1)}=\frac{\lambda_0}{\pi a}
\int_0^\infty \frac{dx}x\frac1{e^{2\pi x/\alpha}-1}\left(\sin^2 x+\frac{x^2}3
\right)=\frac{\lambda_0}{4\pi a}\left[\ln\left(\frac{\sinh\alpha}\alpha\right)
+\frac{\alpha^2}{18}\right],
\label{orderlambda}
\ee
found there both using the Abel-Plana (real frequency) and the Euclidean 
frequency formulations.
 
%%%%%%%%%%%%%%%%%%%%%%%%%%%%%%%%%%%%%%%%%%
\section{Low temperature
\label{lowtsec}}
\subsection{Euclidean frequency argument}\label{ssa}
Let us first write the subtracted free energy in the original point-splitting form:
\be
\Delta F_H=\frac{T}2\sum_{n=-\infty}^\infty e^{i x \tilde\tau}\sum_{l=1}^\infty (2l+1)
P_l(\cos\delta)\ln[x^2-\lambda_0f_H(l,x)],\label{euclidean}
\ee
Here $x=2\pi n a T=n\alpha$.  So the low temperature limit corresponds to small $x$.  Using
the small-argument expansion for the Bessel functions,
\bea
f_H(l,x)&\sim&-\frac{l(l+1)}{2l+1}-\frac{3+2l(l+1)}
{(4l^2-1)(2l+3)}x^2+O(x^4)-x^{2l+1}\left[(-1)^l2^{-2(l+1)}
\frac{(l+1)^2\pi}{\Gamma(l+3/2)^2}+O(x^2)\right], \nn\\
&&\quad x\ll 1,\label{smallx}
\eea
%
%\be
%f_H(l,x)=xe_l'(x)s_l'(x)\sim -\frac{l(l+1)}{2l+1}-\frac{3+2l(l+1)}{(4l^2-1)(2l+3)}x^2+
%O(x^4 \,\mbox{or}\, x^{2l+1}), \quad x\ll1,
%\ee
so it is seen that the leading odd term in $x$ arises only from the $l=1$ term, where
\be
f_H(1,x)\sim -\frac23-\frac7{15}x^2+\frac49 x^3+O(x^4),\quad x\ll1,
\ee
so the logarithm in the free energy is
\be
\ln[x^2-\lambda_0 f_H(1,x)]\sim \ln\frac{2\lambda_0}3+
\left(\frac{3}{2\lambda_0}+\frac7{10}\right)x^2
-\frac23x^3+O(x^4), \quad x\ll1.
\ee
This is the same as Eq.~(6.12) of Ref.~\cite{milton3}, except the $x^2$ in the logarithm
there has been removed by the subtraction.

The above analysis is relevant to the low temperature behavior because that may be
 extracted by using the Euler-Maclaurin formula,
\be
F_H=T\sum_{n=0}^{\infty}{}'g(n)\sim T\int_0^\infty dn\,g(n)-T\sum_{k=1}^\infty 
\frac{B_{2k}}{(2k)!}g^{(2k-1)}(0).\label{em}
\ee
Because of the subtraction, the expansion can be carried out around $n=0$, since the 
function is now analytic there. (In Ref.~\cite{milton3} we did the expansion around 
$n=1$, and we did, in fact, remove the $F_H^{\rm sub}$ term, Eq.~(\ref{fhsub}).
See Eq.~(6.11) there.)
 The integral term  in Eq.~(\ref{em}) is independent
of $T$, %so will be disregarded.  
so the leading contribution to the entropy  comes from the third derivative
term, allowing us to immediately obtain, as before,
\be
\Delta F_H=-\frac2{15}(\pi a)^3T^4,\quad aT\ll \sqrt{\lambda_0},1.\label{stronglow}
\ee
This is the well known strong-coupling low-temperature limit \cite{bd,milton3}.

The above, of course, corresponds to a positive entropy.  But this analysis presumed
that $aT$ was the smallest scale in the problem.  However, we have another parameter,
$\xi=\alpha\sqrt{\frac3{2\lambda_0}}$, which could be large if $\lambda_0\ll\alpha^2$.
The analysis given in Ref.~\cite{milton3} is unchanged, and results in the formula
\be
\Delta F_H  =\left(\frac{2\lambda_0}{3}\right)^2\frac1{\pi a}\left[\frac{\xi^2}{12}
-\ln\xi-\Re\psi\left(1+\frac{i}\xi\right)\right], \quad \alpha\ll1, \xi\sim 1.\label{fhofxi}
\ee
Here $\psi$ is the digamma function. (An alternative derivation is given in Appendix A
of Ref.~\cite{milton4}.)  This function is plotted in Fig.~3 of Ref.~\cite{milton3} and
Fig.~1 of Ref.~\cite{milton4}.  See Fig.~\ref{figfhofxi}  here.
Evidently, the entropy, the negative derivative of
the free energy with respect to temperature, goes negative for sufficiently weak
coupling (large $\xi$), as is seen from the analytic limiting behavior:
\be
\xi\gg 1:\quad \Delta F_H\sim \frac29\lambda_0\pi a T^2, \quad \sqrt{\lambda_0}\ll
 aT\ll 1.\label{lowlow}
\ee
The TE contribution to the entropy is always negative, so the total entropy turns
negative for sufficiently small coupling. 

\subsection{Abel-Plana analysis}
\label{sec:ap}
The derivation of the same result must be achievable directly from the Abel-Plana
form (\ref{fhap}), since the Euler-Maclaurin formula is derivable from the Abel-Plana
expression.  It is a bit subtle, because we have to worry about the appropriate
branch of the phase, but actually very simple.

\begin{figure}
%\centering
\includegraphics{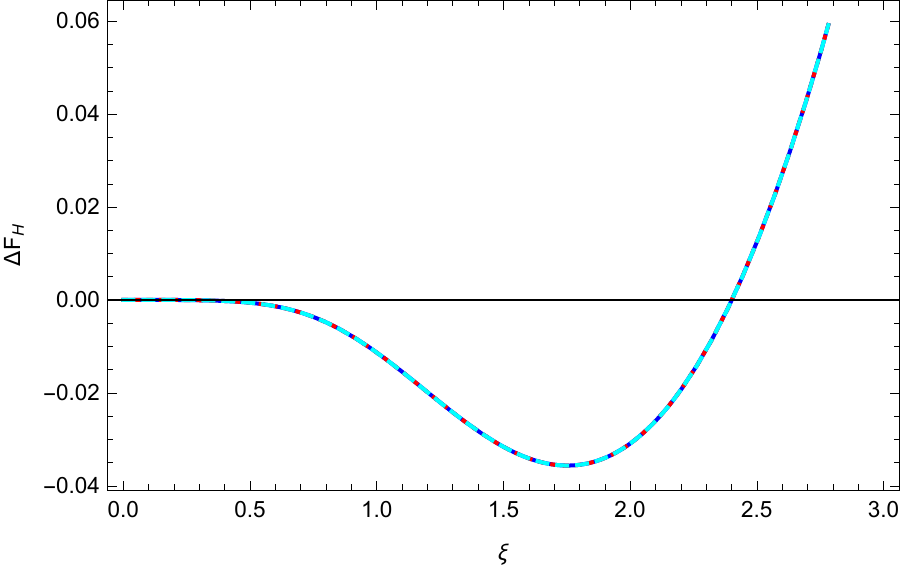}
\caption{\label{figfhofxi} 
The TM free energy for low temperature in terms of
$\xi=\alpha\sqrt{3/(2\lambda_0)}$.  Shown are the coincident results for the
formula (\ref{fhofxi}), and for Eq.~(\ref{fhofxiap}) 
for $\alpha=0.1$ and $\alpha=0.01$. 
Plotted is the free energy apart from a factor of $(2\lambda_0/3)^2/(\pi a)$.
Although the slope is negative (positive entropy) for small $\xi$ (strong coupling),
it is positive (negative entropy) for large enough $\xi$ (weak enough coupling).}
\end{figure}

First, we use  Eq.~(\ref{smallx}) with the replacement $x$ by $ix$.  
(Again, the leading
odd term comes from $l=1$.)  This gives the predominant term in the phase,
($x\ll1$, $x^2/\lambda_0\sim 1$) 
\be
\arg\left[\frac23\lambda_0-\left(1+\frac7{15}\lambda_0\right)x^2+i\frac49\lambda_0x^3
\right]=\arctan\left[\frac{\frac23x^3}{1-\frac{3x^2}{2\lambda_0}}\right].
\ee
The TM free energy thus reads for low $T$
\begin{subequations}
\bea
\Delta F_H&=&-\left(\frac{2\lambda_0}{3}\right)^2\frac1{\pi a}
\frac{3\xi^3}{\alpha^3}
\int_0^\infty dz\frac1{e^{2\pi z/\xi}-1}\arctan\left[\frac{\frac23\left(\frac\alpha
\xi\right)^3z^3}{1-z^2}\right]\label{fhofxiap}\\
&\to& -\left(\frac{2\lambda_0}{3}\right)^2\frac2{\pi a}P\int_0^\infty dz
\frac1{e^{2\pi z/\xi}-1}\frac{z^3}{1-z^2},\quad \alpha\ll 1.\label{fhofxiap2}
\eea
\end{subequations}
These expressions require some explanation.  For the first line,
we remind the reader that, because of our choice of the branch of the arctangent to
be the usual one, %which is continuous at the origin, and $-\pi/2<\arctan(x)\le\pi/2$.  
%This is necessary to have a well-defined limit at zero temperature.  So this means 
there is a discontinuity in the integrand at $z=1$, but of course this is integrable.
We need, for stability, to evaluate the integral by taking a principal value there.
In the second line, we have replaced $\arctan y$ by $y$, appropriate for small $\alpha$,
and the resulting singularity at $z=1$ is integrated by taking a principal value.
Then, numerically, both forms  exactly agree with the previous formula 
(\ref{fhofxi}),  as Fig.~\ref{figfhofxi} shows.
The figure shows that for sufficiently weak coupling, the low-temperature  entropy
turns negative.

It is very easy (much easier than in Sec.~\ref{ssa}) to extract the weak-coupling limit
at low temperature, $\xi\to\infty$.  The crucial observation is that (\ref{fhofxiap2})
receives contributions from only large $z\sim \xi$ when the latter is large, 
so the last factor in the integrand is merely $-z$
%arctangent is well approximated by $-\frac23(\alpha/\xi)^3z$, 
and then the
integral gives the result (\ref{lowlow}) immediately.  Note that the
oddness of the arctangent around $z=0$ is crucial here; were there a
discontinuity in the argument function at $z=0$, the $T\to0$ limit would 
not exist.  
%%%%%%%%%%%%%%%%%%%%%%%%%%%%%%%%%%%%%%%%%%
\section{High temperature}
\label{htempsec}
We showed in Refs.~\cite{milton3,milton4} that the leading behaviors for high temperature
of the TM free energy and entropy are
\be
F_H\sim \frac{\lambda_0}{18}\pi a T^2,\quad S_H=-\frac\partial{\partial T}F_H\sim 
-\frac{\lambda_0}{18}\alpha, \quad \alpha=2\pi a T\gg 1,\lambda_0.\label{high}
\ee
Again, it is remarkable that this is first-order in the coupling.  This same behavior 
was found in Ref.~\cite{bordagandkirsten}.  (If $\lambda_0\gg2\pi a T\gg1$, the
entropy becomes positive \cite{bd}.)
Here, we have made the universal subtraction of the term
$F_H^{\rm sub}$, but that should not alter the conclusion, because that contribution
to the entropy is subdominant at high temperature.  (Indeed, we dropped coupling-independent
terms in Ref.~\cite{milton4}.)

In Ref.~\cite{milton4} we worked out the leading high-temperature form for the free energy
starting from the Euclidean frequency expression (\ref{ftme}) using the uniform asymptotic
expansions for the Riccati-Bessel functions and the Chowla-Selberg formula.  Here, it
seems to be much harder to use the uniform asymptotics on the highly oscillatory real-frequency 
Bessel functions appearing in the Abel-Plana expressions.

\section{Numerical analysis}
\label{numsec}
In principle, it seems that the Abel-Plana formula (\ref{fhap}), which is finite, should be
directly evaluated to obtain the free energy for any temperature and coupling strength.
(It is not possible to do so starting from the Euclidean form (\ref{euclidean}), 
because this still contains divergences.)
The difficulty is that the phase (\ref{argat}) becomes an extremely oscillatory function
for $x>\nu$.  Nevertheless, the sum and integral can be carried out for intermediate
values of $\lambda_0$ and $T$ with moderate computing resources.  %The behavior for 
%large and small values of these parameters is more challenging.

In the numerical calculations, the behaviors of the phase in the vacinity of the
singularities have to be carefully considered.  When the coupling $\lambda_0$ is small,
contributions to the free energy
 near these singularities are significant.  Here, we have carried out the
evaluations with sufficient precision to achieve reliable results, 
limited only by available hardware.

\begin{figure}
%\centering
\begin{subfigure}{.5\textwidth}
\includegraphics[width=\linewidth]{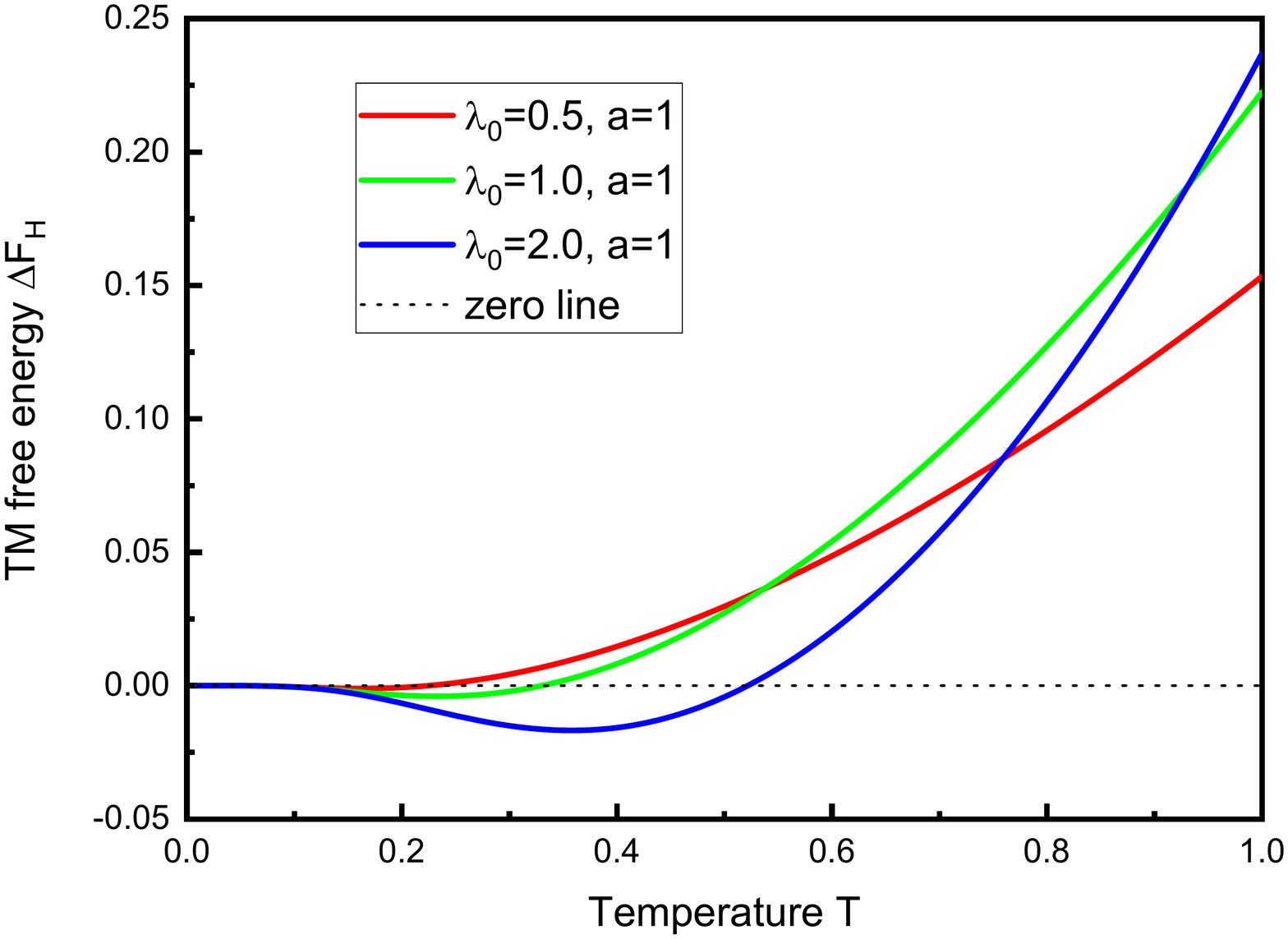}
\caption{Exact free energy.}
%\caption{Temperature behavior of the TM free energy for various intermediate values of $\lambda_0$.}
\label{numfig}
% as a function of temperature (both 
%in units of $1/a$), for different intermediate value of $\lambda_0$.}
\end{subfigure}%
\begin{subfigure}{.5\textwidth}
%\centering
\includegraphics[width=.9\linewidth]{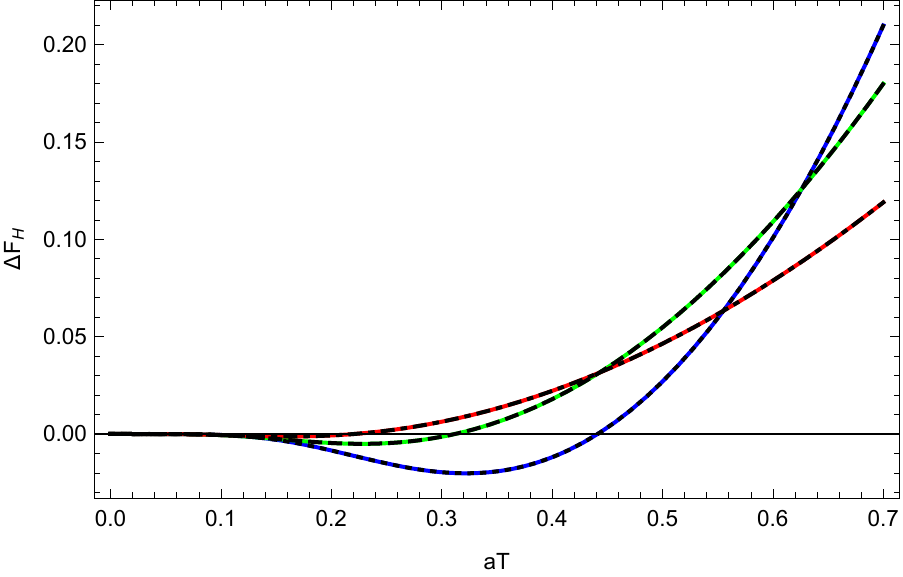}
\caption{Low temperature free energy.}
\label{lowtwrong} 
\end{subfigure}
\caption{ The TM free energy (in units of $1/a$) 
computed from the exact formula (\ref{fhap}) (left panel), or the
low-temperature formula (\ref{fhofxi}) or (\ref{fhofxiap2}) (right panel) plotted as a function of
$aT$ for the same intermediate values of $\lambda_0$,
$\lambda_0=0.5$, 1, and 2, in increasing order on the right side of each figure.  Although 
the low-temperature formula would  not seem to be  applicable here, since the temperature is not 
particularly low, it 
gives results which are qualitatively identical to the exact free energy  seen in 
Fig.~\ref{numfig}, with significant deviations apparent only at higher $T$.}
\label{lowtemp:fig}
\end{figure}

Fig.~\ref{numfig} shows the TM free energy for different moderate values of $\lambda_0$, 
as a function of temperature.
What is truly remarkable is how similar these curves are to those given by the low-temperature
formula (\ref{fhofxi}) which, despite its apparent inappliability, is shown in 
Fig.~\ref{lowtwrong}.
Apparently, then, the numerical results shown in Fig.~\ref{numfig} still largely
inhabit the low-temperature regime.  This is not, perhaps, so surprising, since the validity
of the replacement in Eq.~(\ref{fhofxiap2}) demands $aT\ll1$, not $\alpha\ll1$.

\begin{figure}
%\centering
\begin{subfigure}{.5\textwidth}
%\centering
\includegraphics[width=\linewidth]{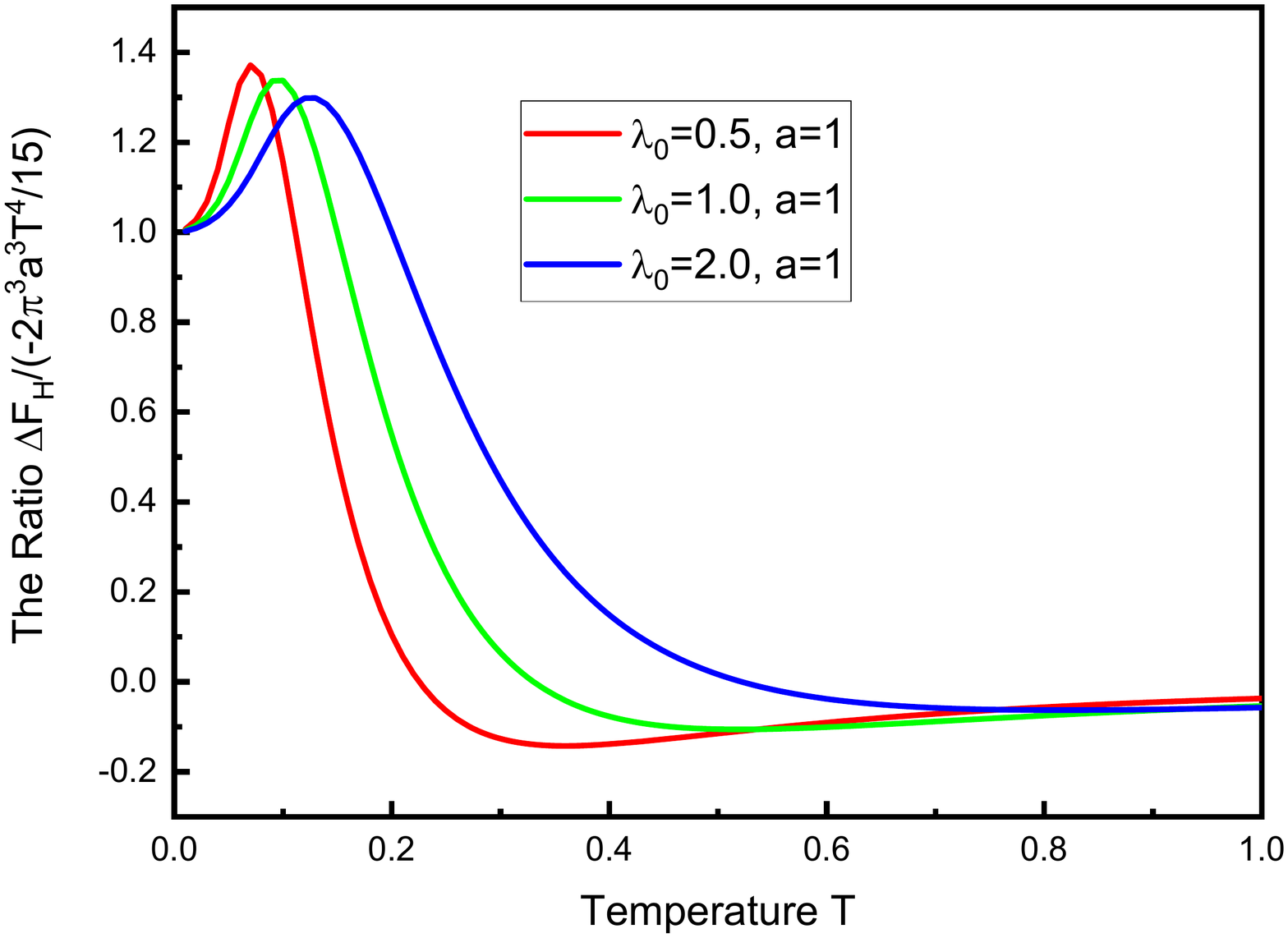}
\caption{Exact free energy}
\label{lowtsc:fig}
\end{subfigure}%
\begin{subfigure}{.5\textwidth}
%\centering
\includegraphics[width=.9\linewidth]{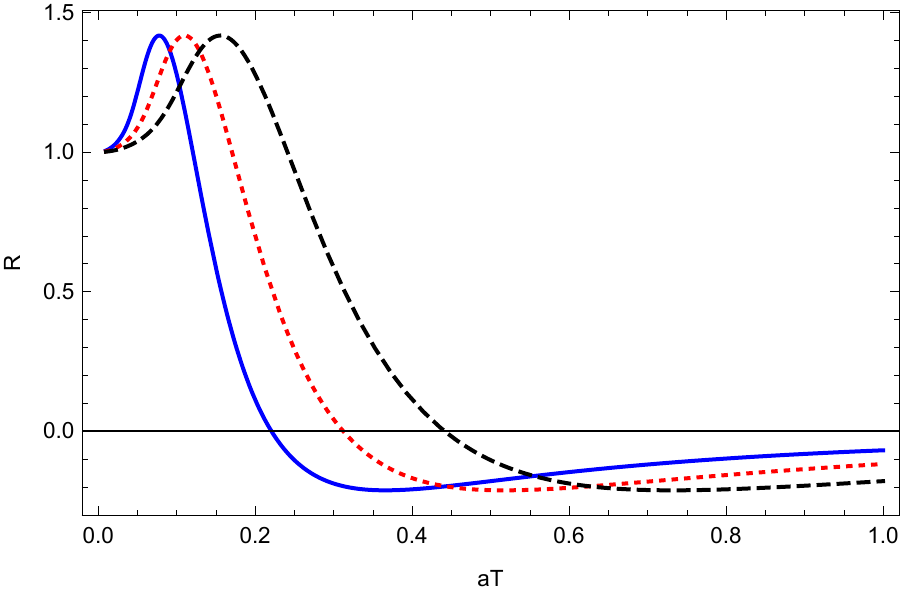}
\caption{Low temperature free energy}
\label{lowta:fig} 
\end{subfigure}
\caption{TM free energy relative to the strong-coupling low-temperature limit.
The left panel shows the exact TM free energy as a function of  temperature $T$ (in
units of $1/a$)  relative to the strong-coupling low-temperature limit (\ref{stronglow}),
for various values of the coupling $\lambda_0$.  For very low temperature,
the free energy agrees with the limit (\ref{stronglow}).  The nonmonotonicity is quite
striking. The right panel shows the ratio $R$ of Eq.~(\ref{fhofxi}) to 
(\ref{stronglow}) as a function of $aT$.  
It is seen that the general low-temperature expression (\ref{fhofxi})
captures most of the behavior shown in Fig.~\ref{lowtsc:fig}. The different curves in 
Fig.~\ref{lowta:fig} correspond to the same values of the coupling as in Fig.~\ref{lowtsc:fig},
namely: $\lambda_0=0.5$ (blue, solid), $\lambda_0=1$
(red, dotted), $\lambda_0=2$ (black, dashed).}
\label{sclt:fig}
\end{figure}

In Fig.~\ref{lowtsc:fig} we compare the computed TM free energy to the strong-coupling
low-temperature result (\ref{stronglow}). This is qualitatively very  similar to that
obtained by taking the ratio of Eqs.~(\ref{fhofxi}) and (\ref{stronglow}), as seen in
Fig.~\ref{lowta:fig}.  Again, this demonstrates that the low temperature description
extends to quite large temperatures.  To put this into perspective, it might help
to note that $aT=1$ corresponds, at room temperature, to a sphere radius of $a=8\,\mu$m.

\begin{figure}
%\centering
\begin{subfigure}{.5\textwidth}
%\centering
\includegraphics[width=\linewidth]{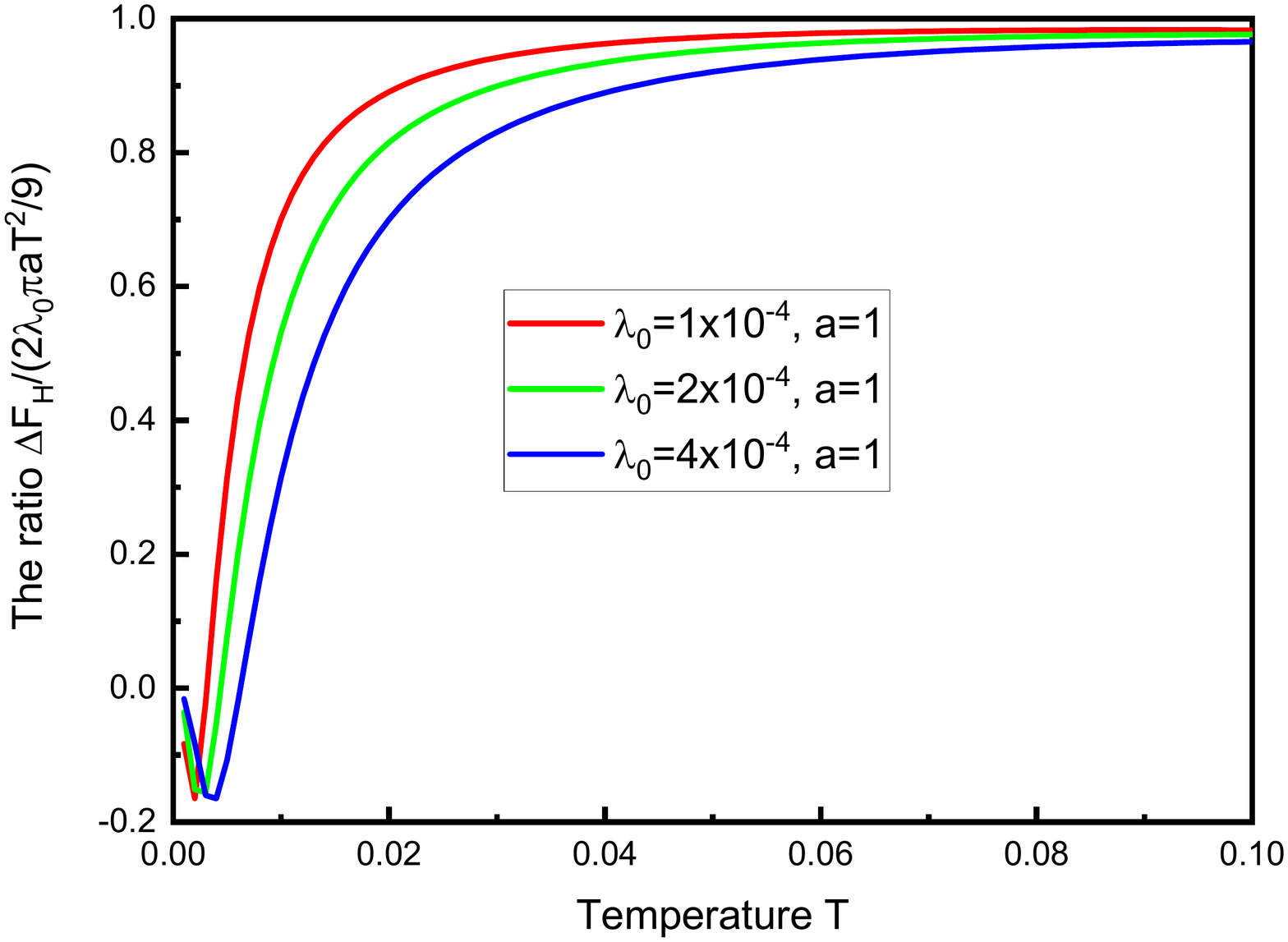}
\caption{Exact free energy}
\label{weaklow:fig}
\end{subfigure}%
\begin{subfigure}{.5\textwidth}
%\centering
\includegraphics[width=.9\linewidth]{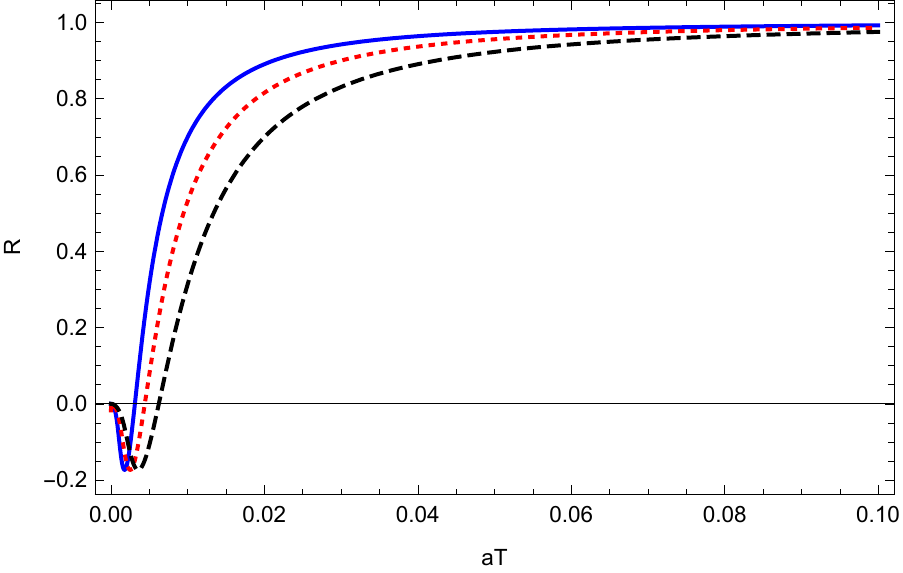}
\caption{Low-temperature free energy}
\label{weaklow2:fig} 
\end{subfigure}
\caption{The  behavior of the TM free energy 
for low temperatures (in units of $1/a$), for even smaller values
of the coupling, relative to the limiting value for low temperature and very small $\lambda_0$,
Eq.~(\ref{lowlow}). The left panel shows the
exact free energy, while the  same ratio $R$
is plotted in the right panel,  except that the
TM free energy is computed from the general low temperature expression (\ref{fhofxi}).  
The different curves are for the same values of $\lambda_0$ as in Fig.~\ref{weaklow:fig}:
$\lambda_0=10^{-4}$ (blue, solid), $\lambda_0=2\times 10^{-4}$ (red, dotted), $\lambda_0
=4 \times 10^{-4}$ (black, dashed). The fact that $F_H$ turns negative for very small temperatures
reflects the limit (\ref{stronglow}).}
\label{wc:fig}
\end{figure}

\begin{figure}
%\centering
\begin{subfigure}{.5\textwidth}
%\centering
\includegraphics[width=\linewidth]{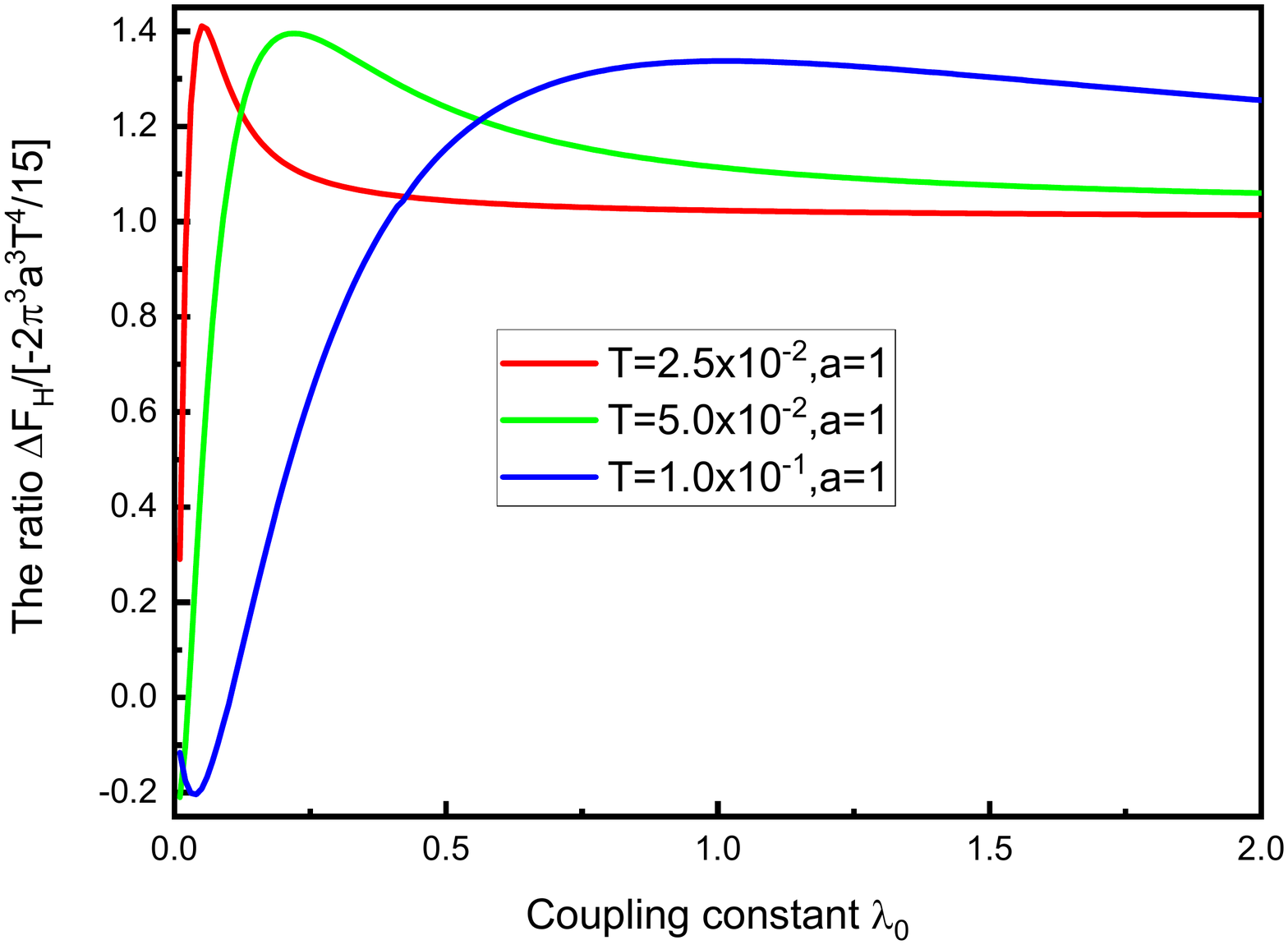}
\caption{Exact free energy}
\label{lowtmpmodc}
\end{subfigure}%
\begin{subfigure}{.5\textwidth}
%\centering
\includegraphics[width=.9\linewidth]{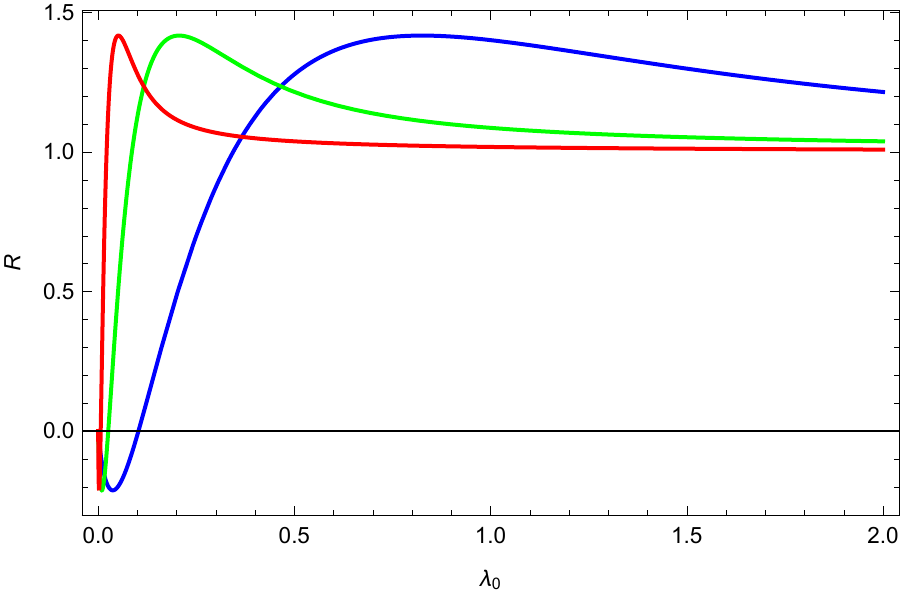}
\caption{Low-temperature free energy}
\label{lowtmodcltf}
\end{subfigure}
\caption{The left panel shows the ratio of the exact TM free energy to the strong-coupling,
low-temperature limit (\ref{stronglow}) for relatively low temperatures, as a function of $\lambda_0$.
The reversal of sign for low $\lambda_0$ reflects the transition from the 
regime where Eq.~(\ref{lowlow})
applies to the strong-coupling, low-temperature limit (\ref{stronglow}).
The right panel shows the same ratio, except instead of the exact free energy, the general 
low-temperature expression (\ref{fhofxi}) is used, for the same values of temperature.
The two graphs are nearly indistinguishable.  In both panels, the different curves 
correspond to the  temperatures $aT=2.5\times 10^{-2}$, $5\times 10^{-2}$, $1\times 10^{-1}$, from
bottom to top on the right of each panel.}
\label{fig:lowmod}
\end{figure}

The weak-coupling regime for low temperature is explored in Fig.~\ref{weaklow:fig}.  The 
comparison here is with Eq.~(\ref{lowlow}).  Of course, this agrees with that obtained
from (\ref{fhofxi}), as demonstrated in Fig.~\ref{weaklow2:fig}.
The low-temperature regime for moderate couplings is explored in Fig.~\ref{lowtmpmodc}.
Again, this agrees with the low-temperature free energy (\ref{fhofxi}), as shown in 
Fig.~\ref{lowtmodcltf}. 

\begin{figure}
%\centering
\includegraphics[width=10cm]{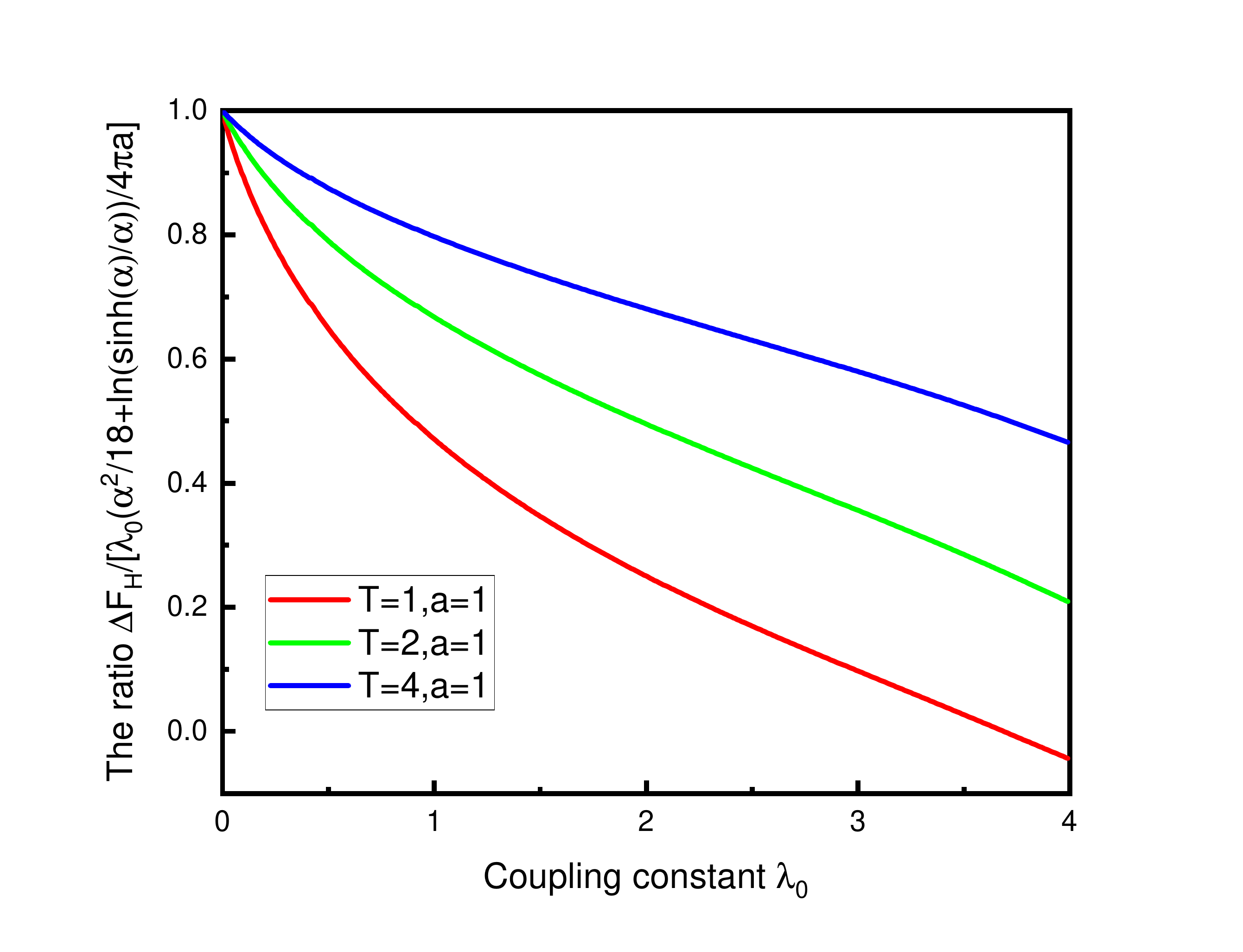}
\caption{\label{highert:fig} The free energy (\ref{fhap})
compared to the $O(\lambda_0)$ approximation
(\ref{orderlambda}).  For small coupling, the ratio approaches unity, and the curves become
flatter for increasing temperature, consistent with the limiting form (\ref{high}).}
\end{figure}

Finally, we compare in Fig.~\ref{highert:fig}
the exact free energy relative to  Eq.~(\ref{orderlambda}).  
We see that the weak-coupling formula is recovered as the coupling goes to zero, and that
the ratio tends to one as the temperature increases, consistent with Eq.~(\ref{high}).

\section{Conclusions}
\begin{figure}
%\centering
\includegraphics{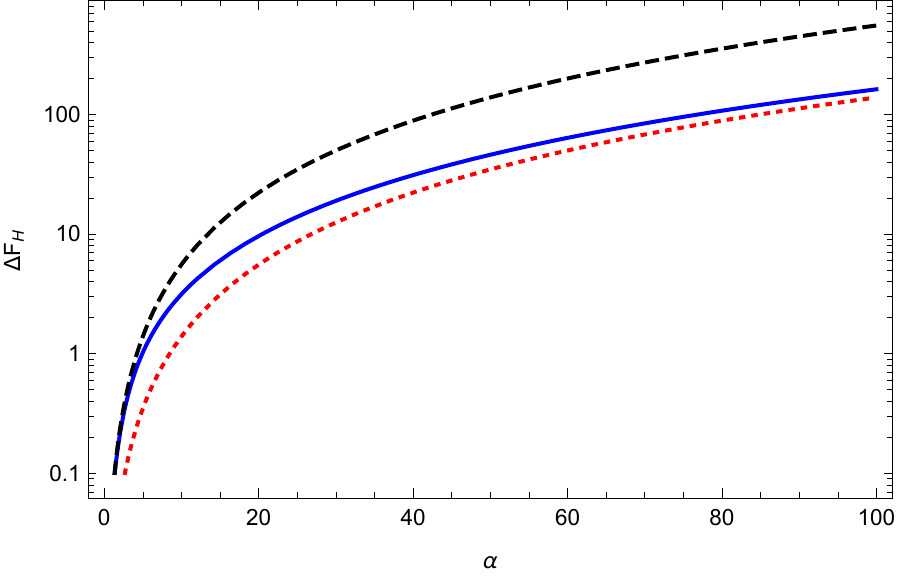}
\caption{\label{fig:weak} The $O(\lambda_0)$ contribution to the free energy (blue, solid), 
given by Eq.~(\ref{orderlambda}), with the prefactor $\lambda_0/(\pi a)$ 
pulled out, 
compared to the limiting forms (\ref{lowlow}) (low temperature, black, dashed)
and (\ref{high}) (high temperature, red, dotted).}
\end{figure}

In this paper, we have re-examined the question of negative entropy for a spherical
 plasma-shell.  We confirm the results first found in Ref.~\cite{milton3}, using now
a uniform subtraction of an irrelevant (infrared) divergent term, 
basing our re-analysis largely on
the Abel-Plana representation of the free energy.  Most interesting is that the leading 
anomalous terms (those corresponding  negative entropy) are captured by the weak-coupling
limit, which we also rederive here.  In Fig.~\ref{fig:weak} we show the weak coupling TM free energy (\ref{orderlambda}) 
compared to the low and high temperature limits, given in Eqs.~(\ref{lowlow}) 
and (\ref{high}), respectively.  
The weak-coupling contribution to the entropy is always negative.

Since the anomalous behavior seems concentrated in the $O(\lambda_0)$ term,
one might be tempted to argue it should be subtracted from the free energy \cite{Graham:2003ib}.
After all, at zero temperature, such terms are frequently recognized as
``tadpole'' terms and are often omitted as unphysical. And for a dielectric ball, at zero 
temperature, the ``bulk subtraction'' also removes automatically the linear term 
in $(\varepsilon-1)$ 
\cite{Milton:2019giz}.  Here, however,
such a subtraction %is impossible.  Doing so 
would ruin the limit to strong
coupling, which has been understood for many years \cite{bd}; see,
for example, Eq.~(\ref{stronglow}).  The
analytic structure of the theory in the coupling constant is rather rigid, 
so {\it ad hoc\/} subtractions are not allowed.  This point was made at the
end of Ref.~\cite{milton3}.
%This section is not mandatory, but can be added to the manuscript if the discussion is unusually long or complex.

In any event, the anomalous behavior is not confined to weak coupling, as the numerical
analysis summarized in Sec.~\ref{numsec} shows.  Therefore, %it is impossible to refute 
the occurence of negative entropy here is hard to deny.  
These remarkable findings may have profound 
implications for our understanding of statistical mechanics and quantum field theory.
\acknowledgments{We thank Gerard Kennedy, Steve Fulling, and Michael Guo for collaborative
assistance. This research was funded by the U.S. National Science Foundation,
 grant numbers 1707511, 2008417.}

\begin{thebibliography}{999}
% Reference 1
\bibitem{schrodinger}
Schr\"odinger, E. {\em What is Life--the Physical Aspect of the Living Cell}.
Cambridge University Press, 1944.

%\bibitem[Author2(year)]{ref-book}
%Author2, L. The title of the cited contribution. In {\em The Book Title}; Editor1, F., Editor2, A., Eds.; Publishing House: City, Country, 2007; pp. 32--58.


%\cite{Cvetic:2001bk}                                                                               
\bibitem{Cvetic:2001bk}
  Cvetic, M., Nojiri, S.,  and Odintsov, S.D.
Black hole thermodynamics and negative entropy in de Sitter and anti-de
Sitter Einstein-Gauss-Bonnet gravity.
{\em  Nucl.\ Phys.\ B} {\bf 2002}, {\em 628}, 295--330.
%  doi:10.1016/S0550-3213(02)00075-5
 % [hep-th/0112045].
  %%CITATION = doi:10.1016/S0550-3213(02)00075-5;%%
  %317 citations counted in INSPIRE as of 2
  % \item 5 Jun 2017                                                

%\cite{Nojiri:2004pf} 
\bibitem{Nojiri:2004pf}
  Nojiri, S.  and Odintsov, S.D.
  The Final state and thermodynamics of dark energy universe. 
{\em  Phys.\ Rev.\ D} {\bf 2004}, {\em 70}, 103522. 
%  doi:10.1103/PhysRevD.70.103522
%  [hep-th/0408170].
  %%CITATION = doi:10.1103/PhysRevD.70.103522;%% 
  %433 citations counted in INSPIRE as of 12 Jun 2017     


\bibitem[Brevik(2006)]{njp}
  Brevik, I.,   Ellingsen, S.A.,  and Milton, K.A.
Thermal corrections to the Casimir effect.
{\em New J. Phys.} {\bf 2006}, {\em 8}, 236.  %[quant-ph/0605005].


\bibitem[Bezerra(2008)]{bezerra}
Bezerra, V.B.,  Klimchitskaya, G.L., Mostepanenko, V. M., and Romero, C.
Lifshitz theory of atom-wall interaction with applications to quantum
reflection. {\em Phys.\ Rev.\ A} {\bf2008}, {\em 78}, 042901.

\bibitem[Canaguier(2010)]{durand1}
Canaguier-Durand, A., Maia Neto, P. A., Lambrecht, A., and  Reynaud, S.
Thermal Casimir effect in the plane-sphere geometry.
{\em Phys.\ Rev.\ Lett.\ } {\bf2010}, {\em104}, 040403.

\bibitem[Canaguier(2010a)]{durand2}
Canaguier-Durand, A.,  Maia Neto, P.A., Lambrecht, A., and Reynaud, S.
Thermal Casimir effect for Drude metals in the plane-sphere geometry.
{\em Phys.\ Rev.\ A} {\bf2010}, {\em 82}, 012511.

\bibitem[Bordag(2010)]{bp}
  Bordag, M. and Pirozhenko, I.G.,
  Casimir entropy for a ball in front of a plane.
{\em  Phys.\ Rev.\ D} {\bf2010}, {\em 82}, 125016.
%  doi:10.1103/PhysRevD.82.125016
 % [arXiv:1010.1217 [quant-ph]].
  %%CITATION = doi:10.1103/PhysRevD.82.125016;%%                                      
  %14 citations counted in INSPIRE as of 27 Jul 2017                                  

\bibitem[Rodriguez(2011)]{lopez1}
 Rodriguez-Lopez, P.
Casimir energy and entropy in the sphere--sphere geometry.
{\em Phys.\ Rev.\ B} {\bf 2011}, {\em 84}, 075431.

\bibitem[Rodriguez(2012)]{lopez2}
 Rodriguez-Lopez, P.
Casimir energy and entropy between perfect metal spheres. In
{\em Quantum Field Theory Under the Influence of External Conditions (QFEXT11)},
{\em Int.\ J. Mod.\ Phys.: Conf.\ Ser.} {\bf2012}, {\em 14}, 475-484.

\bibitem[Milton(2015)]{milton}
Milton, K.A.,  Gu\'erout, R.,   Ingold, G.-L., Lambrecht, A., and   Reynaud, S.
Negative Casimir entropies in nanoparticle interactions.
{\em J. Phys.: Condens.\ Matter} {\bf2015}, {\em27}, 214003.

\bibitem[Ingold(2015)]{ingold}
Ingold, G.-L.,  Umrath, S., Hartmann, M.,  Gu\'erout,  R.,  Lambrecht, A.,
Reynaud, S.,  and Milton, K.A.
Geometric origin of negative Casimir entropies: A scattering-channel analysis.
{\em Phys.\ Rev.\ E} {\bf2015}, {\em91}, 033203.



\bibitem[Li(2016)]{li}
Li, Y.,   Milton, K.A., Kalauni, P.,  and  Parashar, P. Casimir self-%          
entropy of an electromagnetic thin sheet. {\em Phys.\ Rev.\ D}  {\bf2016}, {\em94}, 
085010.

\bibitem[Milton(2017)]{milton2}
Milton, K.A.,  Li, Y.,  Kalauni, P.,  Parashar, P., Gu\'erout, R., 
Ingold, G.-L.,  Lambrecht, A., and  Reynaud, S.
Negative entropies in Casimir and Casimir-Polder interactions.
%extended version of talk at Frontiers of Quantum and Mesoscopic Thermodynamics, 
%Prague, 2015,     
{\em Fortschr.\ Phys.} {\bf2017}, {\em65}, 1600047. % DOI: 10.1002/prop.201600047
%[arXiv:1605.01073].

\bibitem[Balian(1978)]{bd}
 Balian, R. and Duplantier, B.
  Electromagnetic waves near perfect conductors. 2. Casimir effect.
{\em  Ann.\ Phys.\ (N.Y.)} {\bf1978}, {\em 112}, 165--208.
%  doi:10.1016/0003-4916(78)90083-0
  %%CITATION = doi:10.1016/0003-4916(78)90083-0;%%                                    
  %274 citations counted in INSPIRE as of 26 May 2016                                 

\bibitem[Milton(2017)]{milton3}
 Milton, K.A.,  Kalauni, P.,  Parashar, P.,  and Li, Y. 
Casimir self-entropy of a spherical electromagnetic $\delta$-function shell.
%arXiv: 1707.09840, 
{\em Phys. Rev. D} {\bf2017}, {\em 96}, 085007.

\bibitem[Milton(2019)]{milton4}
 Milton, K.A.,  Kalauni, P.,  Parashar, P., and Li, Y. 
Remarks on the Casimir self-entropy of a spherical electromagnetic $\delta$-function 
shell.
%arXiv:1808.03816, 
{\em Phys. Rev. D} {\bf2019}, {\em 99}, 045013.

\bibitem{bordagandkirsten}
Bordag, M. and Kirsten, K.
On the entropy of a spherical plasma shell.
%arXiv:1805.11241.
{\em J. Phys. A} {\bf2018}, {\em 51}, 455001.

\bibitem{bordagfs}
Bordag, M. Free energy and entropy for thin sheets.
{\em Phys.\ Rev.\ D } {\bf2018}, {\em 98}, 085010.
% [arXiv:1807.09691].

\bibitem{deltasph}
 Parashar, P.,   Milton, K.A.,  Shajesh, K.V., and  Brevik, I. 
Electromagnetic $\delta$-function sphere. 
%arXiv:1708.01222, 
{\em Phys. Rev. D} {\bf2017}, {\em 96}, 085010.

  %\cite{Graham:2003ib}                                                                              
\bibitem{Graham:2003ib}
 Graham, N., Jaffe, R.L., Khemani, V., Quandt, M., Schr\"oder, O., and ~Weigel, H.
  The Dirichlet Casimir problem.
 {\em Nucl.\ Phys.\ B} {\bf2004}, {\em 677}, 379-404.
%  doi:10.1016/j.nuclphysb.2003.11.001
 % [hep-th/0309130].
  %%CITATION = doi:10.1016/j.nuclphysb.2003.11.001;%%                                                
  %123 citations counted in INSPIRE as of 14 Jun 2017   


\bibitem{Milton:2019giz}
    Milton, K.A., Parashar, P., Brevik, I., and Kennedy, G.
    Self-stress on a dielectric ball and Casimir\textendash{}Polder forces.
{\em Ann.\ Phys.\ (N.Y.)} {\bf 2020}, {\em 412}, 168008.    
%eprint = ``1909.05721'',
 %   archivePrefix = ``arXiv'',
  %  primaryClass = ``hep-th'',
   % doi = ``10.1016/j.aop.2019.168008'',
    %journal = ``Annals Phys.'',
    %volume = ``412'',
    %pages = ``168008'',
    %year = ``2020''


\end{thebibliography}
\end{document}